\documentclass[aps,prb,superscriptaddress,longbibliography,reprint]{revtex4-1}

\usepackage{amsmath}
\usepackage{amssymb}
\usepackage{bm}
\usepackage[usenames,dvipsnames]{color}
\usepackage{graphicx}
\usepackage[latin1]{inputenc}
\usepackage[english]{babel}
\usepackage{array}

\begin{document}

\title{Tunable damping, saturation magnetization, and exchange stiffness of half-Heusler NiMnSb thin films}

\author{P.~D\"urrenfeld}
\affiliation{Department of Physics, University of Gothenburg, 412 96 Gothenburg, Sweden}

\author{F.~Gerhard}
\affiliation{Physikalisches Institut (EP3), Universit\"at W\"urzburg, 97074 W\"urzburg, Germany}

\author{J.~Chico}
\affiliation{Department of Physics and Astronomy, Uppsala University, Box 520, 752 20 Uppsala, Sweden}

\author{R.~K.~Dumas}
\affiliation{Department of Physics, University of Gothenburg, 412 96 Gothenburg, Sweden}
\affiliation{NanOsc AB, 164 40 Kista, Sweden}

\author{M.~Ranjbar}
\affiliation{Department of Physics, University of Gothenburg, 412 96 Gothenburg, Sweden}

\author{A.~Bergman}
\affiliation{Department of Physics and Astronomy, Uppsala University, Box 520, 752 20 Uppsala, Sweden}

\author{L.~Bergqvist}
\affiliation{Materials and Nano Physics, School of ICT, KTH Royal Institute of Technology, Electrum 229, 164 40 Kista, Sweden}
\affiliation{Swedish e-Science Research Centre (SeRC), 100 44 Stockholm, Sweden}

\author{A.~Delin}
\affiliation{Department of Physics and Astronomy, Uppsala University, Box 520, 752 20 Uppsala, Sweden}
\affiliation{Materials and Nano Physics, School of ICT, KTH Royal Institute of Technology, Electrum 229, 164 40 Kista, Sweden}
\affiliation{Swedish e-Science Research Centre (SeRC), 100 44 Stockholm, Sweden}

\author{C.~Gould}
\affiliation{Physikalisches Institut (EP3), Universit\"at W\"urzburg, 97074 W\"urzburg, Germany}

\author{L.~W.~Molenkamp}
\affiliation{Physikalisches Institut (EP3), Universit\"at W\"urzburg, 97074 W\"urzburg, Germany}

\author{J.~\AA kerman}
\affiliation{Department of Physics, University of Gothenburg, 412 96 Gothenburg, Sweden}
\affiliation{NanOsc AB, 164 40 Kista, Sweden}
\affiliation{Materials and Nano Physics, School of ICT, KTH Royal Institute of Technology, Electrum 229, 164 40 Kista, Sweden}

\begin{abstract}
The half-metallic half-Heusler alloy NiMnSb is a promising candidate for applications in spintronic devices due to its low magnetic damping and its rich anisotropies. Here we use ferromagnetic resonance (FMR) measurements and calculations from first principles to investigate how the composition of the epitaxially grown NiMnSb influences the magnetodynamic properties of saturation magnetization $M_S$, Gilbert damping $\alpha$, and exchange stiffness $A$. $M_S$ and $A$ are shown to have a maximum for stoichiometric composition, while the Gilbert damping is minimum. We find excellent quantitative agreement between theory and experiment for $M_S$ and $\alpha$. The calculated $A$ shows the same trend as the experimental data, but has  a larger magnitude. Additionally to the unique in-plane anisotropy of the material, these tunabilities of the magnetodynamic properties can be taken advantage of when employing NiMnSb films in magnonic devices.
\end{abstract}

\maketitle

\section{Introduction}

Interest in the use of half-metallic Heusler and half-Heusler alloys in spintronic and magnonic devices is steadily increasing,~\cite{okura2011apl,yamamoto2015apl,duerrenfeld2015jap} as these materials typically exhibit both a very high spin polarization~\cite{degroot1983prl,ristoiu2000epl,kubota2009apl,muller2009ntm,jourdan2014ntc} and very low spin-wave damping.~\cite{heinrich2004jap,koveshnikov2005jap,trudel2010jpd,riegler2011thesis} One such material is the epitaxially grown half-Heusler alloy NiMnSb,~\cite{bach2003apl,bach2003jcg} which not only has one of the lowest known spin-wave damping values of any magnetic metal, but also exhibits an interesting and tunable combination of two-fold in-plane anisotropy~\cite{gerhard2014jap} and moderate out-of-plane anisotropy,~\cite{koveshnikov2005jap} all potentially interesting properties for use in both nanocontact-based spin-torque oscillators~\cite{tsoi1998prl,slonczewski1999jmmm,silva2008jmmm,madami2011nn,bonetti2010prl,dumas2013prl,dumas2014ieeem} and spin Hall nano-oscillators~\cite{demidov2012ntm,liu2013prl,demidov2014ntc,duan2014ntc,ranjbar2014ieeeml}. To successfully employ NiMnSb in such devices, it is crucial to understand, control, and tailor both its magnetostatic and magnetodynamic properties, such as its Gilbert damping ($\alpha$), saturation magnetization ($M_S$), and exchange stiffness ($A$).

Here we investigate these properties in Ni$_{\text{1-x}}$Mn$_{\text{1+x}}$Sb films using ferromagnetic resonance (FMR) measurements and calculations from first principles for compositions of -0.1 $\leq \text{x} \leq$ 0.4. $M_S$ and $A$ are shown experimentally to have a maximum for stoichiometric composition, while the Gilbert damping is minimum; this is in excellent quantitative agreement with  calculations of and experiment on $M_S$ and $\alpha$. The calculated $A$ shows the same trend as the experimental data, but with an overall larger magnitude. We also demonstrate that the exchange stiffness can be easily tuned over a wide range in NiMnSb through Mn doping, and that the ultra-low damping persists over a wide range of exchange stiffnesses. This unique behavior makes NiMnSb ideal for tailored spintronic and magnonic devices.
Finally, by comparing the experimental results with first-principles calculations, we also conclude that the excess Mn mainly occupies Ni sites and that interstitial doping plays only a minor role.

\section{Methods}

\begin{table}[b]
\centering
\begin{tabular}{>{\centering\arraybackslash}m{1.1cm}  >{\centering\arraybackslash}m{2cm}  >{\centering\arraybackslash}m{1.6cm}  >{\centering\arraybackslash}m{1.6cm} >{\centering\arraybackslash}m{1.6cm}}
 \hline \hline
Sample & vertical lattice constant (\AA) & thickness (nm) & uniaxial easy axis & $\frac{2 K_{1}}{M_{S}}$ (Oe)\\
 \hline
1 & 5.94 & 38 & [110] & 170\\
2 & 5.97 & 38 & [110] & 8.4\\
3 & 5.99 & 40 & [110] & 0\\
4 & 6.02 & 45 & [1$\bar{{1}}$0] & 9.0\\
5 & 6.06 & 45 & [1$\bar{{1}}$0] & 14.2\\
6 & 6.09 & 38 & [1$\bar{{1}}$0] & 25.5\\
\hline \hline
\end{tabular}
\caption{Overview of NiMnSb films investigated in this study.}
\label{tab:SampleOverview}
\end{table}

\subsection{Thin Film Growth}
The NiMnSb films were grown by molecular beam epitaxy onto InP(001) substrates after deposition of a 200~nm thick (In,Ga)As buffer layer.~\cite{gerhard2014jap} The films were subsequently covered \textit{in situ} by a 10~nm thick magnetron sputtered metal cap to avoid oxidation and surface relaxation.~\cite{kumpf2007pssc} The Mn content was controlled during growth via the temperature, and hence the flux, of the Mn effusion cell. Six different samples (see table~\ref{tab:SampleOverview}) were grown with increasing Mn concentration, sample 1 having the lowest and sample 6 the highest  concentration of Mn. High-resolution x-ray diffraction (HRXRD) measurements give information on the structural properties of these samples, confirming the extremely high crystalline quality of all samples with different Mn concentration, even in the far from stoichiometric cases (samples 1 and 6).~\cite{gerhard2014jap} The vertical lattice constant is found to increase with increasing Mn concentration and, assuming a linear increase,~\cite{ekholm2010jap} we estimate the difference in Mn concentration across the whole set of samples to be about 40~at.~\%. We will thus represent the Mn concentration in the following experimental results by the measured vertical lattice constant. Stoichiometric NiMnSb exhibits vertical lattice constants in the range of 5.96--6.00~\AA, leading to the expectation of stoichiometric NiMnSb in samples 2 and 3.~\cite{gerhard2014jap} Finally, the layer thicknesses are also determined from the HRXRD measurements, giving an accuracy of $\pm$1~nm.

\subsection{Ferromagnetic Resonance}
\begin{figure}[t]
\includegraphics*[width=85mm]{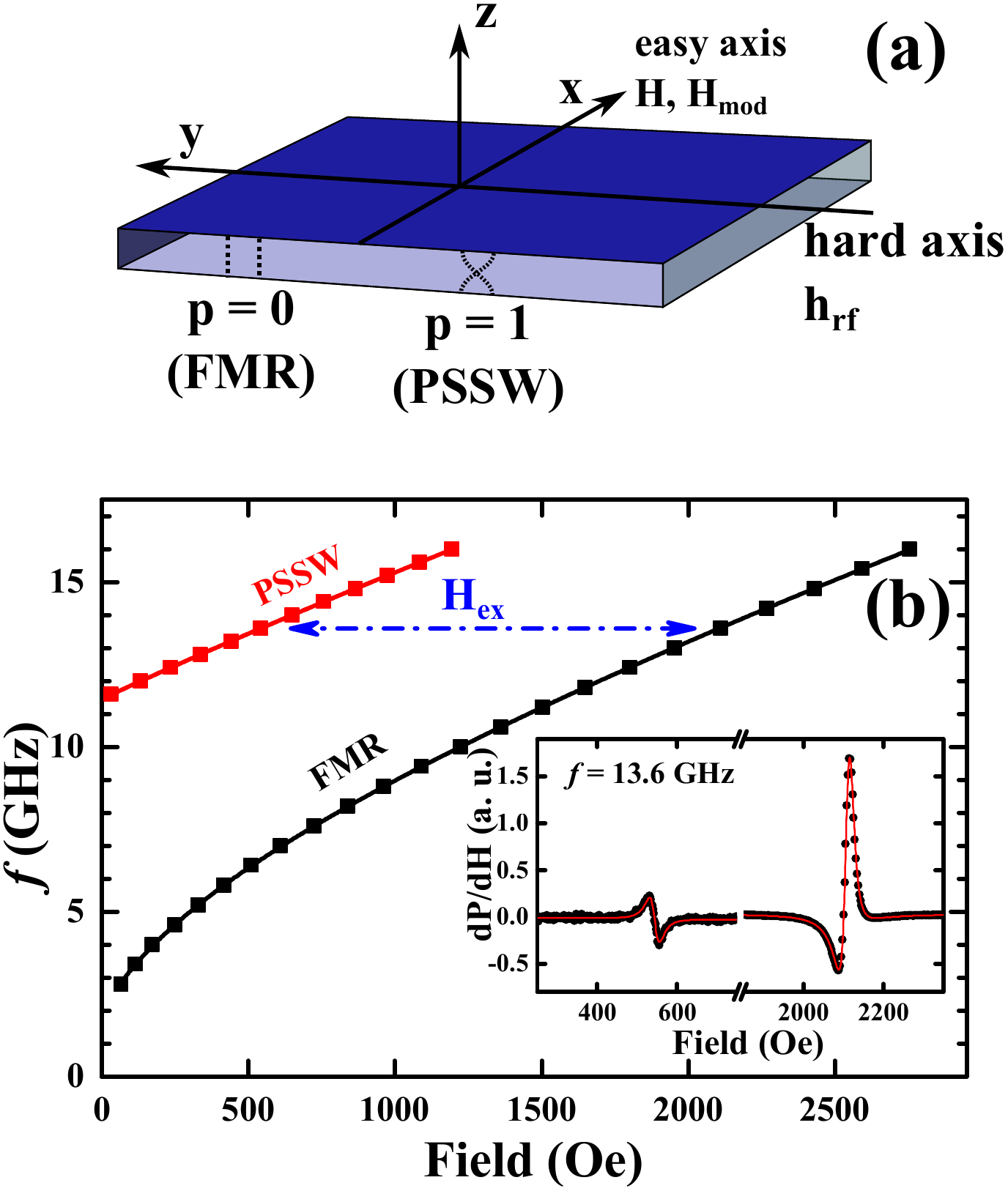}
\caption{(a) Schematic diagram of the FMR measurement showing field directions. In our setup, the FMR mode and the first PSSW mode are excited. (b) Frequency vs. resonance fields of the PSSW (red) and uniform FMR (black) mode for sample 2. The solid lines are fits to the Kittel equation, and both modes are offset horizontally by $H_{ex}$. Inset: Resonance curves for $f$=13.6~GHz. The first PSSW mode on the left and the FMR mode on the right were fit with Eq.~\ref{eq:FMR}}\label{fig:fig1}
\end{figure}

Broadband field-swept FMR spectroscopy was performed using a NanOsc Instruments PhaseFMR system with a coplanar waveguide for microwave field excitation. Microwave fields $h_{rf}$ with frequencies of up to 16~GHz were applied in the film plane, perpendicularly oriented to an in-plane dc magnetic field $H$. The derivative of the FMR absorption signal was measured using a lock-in technique, in which an additional low-frequency modulation field $H_{mod}<$1~Oe was applied using a pair of Helmholtz coils parallel to the dc magnetic field. The field directions are shown schematically in Fig.~\ref{fig:fig1}(a) and a typical spectrum measured at 13.6 GHz is given in the inset of Fig.~\ref{fig:fig1}(b). In addition to the zero wave vector uniform FMR mode seen at about $H$=2.1 kOe, an additional weaker resonance is observed at a much lower field of about 500 Oe, and is identified as the first exchange-dominated perpendicular standing spin wave (PSSW) mode. The PSSW mode has a nonzero wave vector pointing perpendicular to the thin film plane and a thickness-dependent spin-wave amplitude and phase.~\cite{portis1963apl,lenk2010prb} This can be efficiently excited in the coplanar waveguide geometry due to the nonuniform strength of the microwave field across the film thickness.~\cite{maksymov2015pe}

The field dependence of the absorption spectra (inset of Fig.~\ref{fig:fig1}(b)) can be fit well (red line) by the sum of a symmetric and an antisymmetric Lorentzian derivative:~\cite{woltersdorf2004thesis,mecking2007prb}
\begin{eqnarray}
\label{eq:FMR}
\frac{dP}{dH} (H) &=& \frac{-8 C_1 \Delta H (H - H_{0})}{\left[ \Delta H^2 + 4 \left( H - H_{0} \right)^2 \right]^2} \nonumber \\
&& + \frac{2 C_2 \left( \Delta H^2 - 4 (H - H_{0})^2 \right)}{\left[ \Delta H^2 + 4 \left( H - H_{0} \right)^2 \right]^2}{,}
\end{eqnarray}
where $H_{0}$ is the resonance field, $\Delta H$ the full width at half maximum (FWHM), and $C_1$ and $C_2$ fitted parameters representing the amplitude of the symmetric and antisymmetric Lorentzian derivatives, respectively. Both the FMR and the PSSW peaks can be fitted independently, as they are well separated by the exchange field $\mu_{0} H_{ex} \propto (\pi/d)^2$, where $d$ is the thickness of the layer. For our chosen sample thicknesses, the differences in resonance fields are always much larger than the resonance linewidths.

The field dependence of both resonances is shown in Fig.~\ref{fig:fig1}(b) and can now be used to extract information about the magnetodynamic properties and anisotropies of the films. The curves are fits to the Kittel equation, including internal fields from the anisotropy and the exchange field for the PSSW excitation:~\cite{gerhard2014jap,conca2014apl}
\begin{eqnarray}
\label{eq:KittelAnisotropies}
f &=& \frac{\gamma \mu_0}{2 \pi} \bigg[ \left( H_{0} + \frac{2 K_{U}}{M_{S}} - \frac{2 K_{1}}{M_{S}} + H_{ex} \right) \nonumber\\
&& \times \left( H_{0} + \frac{2 K_{U}}{M_{S}} + \frac{K_{1}}{M_{S}} + H_{ex} + M_{\text{eff}} \right) \bigg]^{1/2}{,}
\end{eqnarray}
where $H_{0}$ is the resonance field, $\gamma/2 \pi$  the gyromagnetic ratio, and $\mu_{0}$ the permeability of free space. $M_{\text{eff}}$ is the effective magnetization, which has a value close to the saturation magnetization $M_{S}$. $2 K_{U}/M_{S}$ and $2 K_{1}/M_{S}$ stands for the internal anisotropy fields coming from the uniaxial ($K_{U}$) and biaxial ($K_{1}$) anisotropy energy densities in the half-Heusler material. The effective magnetic field also includes an exchange field $\mu_{0} H_{ex} = (2A/M_{S})(p \pi/d)^2$, which is related to the exchange stiffness $A$, the film thickness $d$, and the integer order of the PSSW mode $p$, where $p$~=~0 denotes the uniform FMR excitation and $p$~=~1 the first PSSW mode. This mode numbering reflects the boundary conditions with no surface pinning of the spins, which is expected for the in-plane measurement geometry.~\cite{vankampen2002prl}

We stress that the expression for the anisotropy contribution in Eq.~\ref{eq:KittelAnisotropies} is only valid for the case in which the magnetization direction is parallel to the uniaxial easy axis and also parallel to the applied field. A full angular-dependent formulation of the FMR condition is described in Ref.~\onlinecite{gerhard2014jap}. To fulfill the condition of parallel alignment for all resonances, we perform the FMR measurements with the dc magnetic field being applied along the dominant uniaxial easy axis of each film, which changes from the [110] crystallographic direction to the [1$\bar{{1}}$0]-direction with increasing Mn concentration (see Table~\ref{tab:SampleOverview}).

The values of the biaxial anisotropy $\frac{2 K_{1}}{M_{S}}$ have been determined in a previous study by fixed-frequency in-plane angular dependent FMR measurements,~\cite{gerhard2014jap} and were thus taken as constant values in the fitting process for  Eq.~\ref{eq:KittelAnisotropies}; a simultaneous fit of both contributions can yield arbitrary combinations of anisotropy fields due to their great interdependence. The values for the uniaxial anisotropy $\frac{2 K_{U}}{M_{S}}$ obtained from the frequency-dependent fitting are in very good agreement with the previously obtained values in Ref.~\onlinecite{gerhard2014jap}. The gyromagnetic ratio was measured to be $\gamma/2 \pi$~=~(28.59$\pm$0.20)~GHz/T for all investigated samples, and was therefore fixed for all samples to allow better comparison of the effective magnetization values.

The Gilbert damping $\alpha$ of the films is obtained by fitting the FMR linewidths $\Delta H$ with the linear dependence:~\cite{kalarickal2006jap}
\begin{equation}
\label{eq:damping}
    \mu_0 \Delta H = \mu_0 \Delta H_{0} + \frac{4 \pi \alpha}{\gamma} f {,}
\end{equation}
where $\Delta H_{0}$ is the inhomogeneous linewidth broadening of the film. The parallel alignment between magnetization and external magnetic field ensures that the linewidth is determined by the Gilbert damping process only.~\cite{zakeri2007prb}

\subsection{Calculations from First Principles}

The electronic and magnetic properties of the NiMnSb half-Heusler system were studied via first-principles calculations. The material was assumed to be ordered in a face-centered tetragonal structure with an in-plane lattice parameter $a_\text{lat}^{\parallel}$~=~5.88~{\AA}, close to the lattice constant of the InP substrate, and an out-of-plane lattice constant of $a_\text{lat}^{\bot}$~=~5.99~\AA, matching the value for the stoichiometric composition. Fixed values for the lattice parameters were chosen since an exact relation between the off-stoichiometric composition and the experimentally measured vertical lattice constants cannot be established. Moreover, calculations with a varying vertical lattice parameter for a constant composition showed only a negligible effect on M$_{S}$, $A$, and $\alpha$. The calculations were performed using the multiple scattering Korringa-Kohn-Rostocker (KKR) Green's function formalism as implemented in the SPRKKR package.~\cite{ebert2011rpp} Relativistic effects were fully taken into account by solving the  Dirac equation for the electronic states, the shape of the potential was considered via the Atomic Sphere Approximation (ASA), and the local spin density approximation (LSDA) was used for the exchange correlation potential. The coherent potential approximation (CPA) was used for the chemical disorder of the system. 

The Gilbert damping $\alpha$ of the material was calculated using linear response theory~\cite{garate2009prb}, including the temperature effects from interatomic displacements and spin fluctuations.~\cite{mankovsky2013prb,ebert2015prb}

The exchange interactions $J_{ij}$ between the atomic magnetic moments were calculated using the magnetic force theorem, as considered in the LKAG formalism.~\cite{liechtenstein1984jpf,liechtenstein1987jmmm} The interactions were calculated for up to 4.5 times the lattice constant in order to take into account any long-range interactions. Given the interatomic exchange interactions, the \textit{spin-wave stiffness} $D$ can be calculated. Due to possible oscillations in the exchange interactions as a function of the distance, it becomes necessary to introduce a damping parameter, $\eta$, to assure convergence of the summation. $D$ can then be obtained by evaluating the limit $\eta\rightarrow 0$ of 
\begin{equation}
  D=\frac{2}{3}\sum_{ij}\frac{J_{ij}}{\sqrt{M_i M_j}}r_{ij}^2\exp\left(-\eta\frac{r_{ij}}{a_\text{lat}}\right),
  \label{eq:exchange_stiff}
 \end{equation}
as described in~[\onlinecite{pajda2001prb}].
Here, $M_i$ and $M_j$ are the local magnetic moments at sites $i$ and $j$, $J_{ij}$ is the exchange coupling between the magnetic moments at sites $i$ and $j$, and $r_{ij}$ is the distance between the atoms $i$ and $j$. This formalism can be extended to a multisublattice system~\cite{thoene2009jpd}.

To calculate the effect of chemical disorder on the exchange stiffness of the system, the obtained exchange interactions were summed over a supercell with a random distribution of atoms in the chemically disordered sublattice. The effect that distinct chemical configurations can have over the calculation of the exchange stiffness was treated by taking 200 different supercells. The results were then averaged and the standard deviation was calculated. The cells were obtained using the atomistic spin dynamics package UppASD.~\cite{skubic2008jpcm}

Finally, with the spin-wave stiffness determined as described above, the \textit{exchange stiffness} $A$ can be calculated from:~\cite{vaz2008rpp}
\begin{equation}
 A=\frac{D M_\text{S}\left(T\right)}{2g\mu_B}.
\end{equation}
Here, $g$ is the Land\'e g-factor of the electron, $\mu_B$ the Bohr magneton, and $M_\text{S}\left(T\right)$  the magnetization density of the system for a given temperature $T$, which for $T=0$\,K corresponds to the saturation magnetization.

From the first-principles calculations, the magnetic properties for ordered NiMnSb and chemically disordered Ni$_{\text{1-x}}$Mn$_{\text{1+x}}$Sb were studied. To obtain the values of the exchange stiffness $A$ for $T=300$\,K, the exchange interactions from the ab initio calculations were used in conjunction with the value of the magnetization at $T=300$\,K obtained from Monte Carlo simulations.

\section{Results}

\subsection{Magnetization}

\begin{figure}[t]
\includegraphics*[width=70mm]{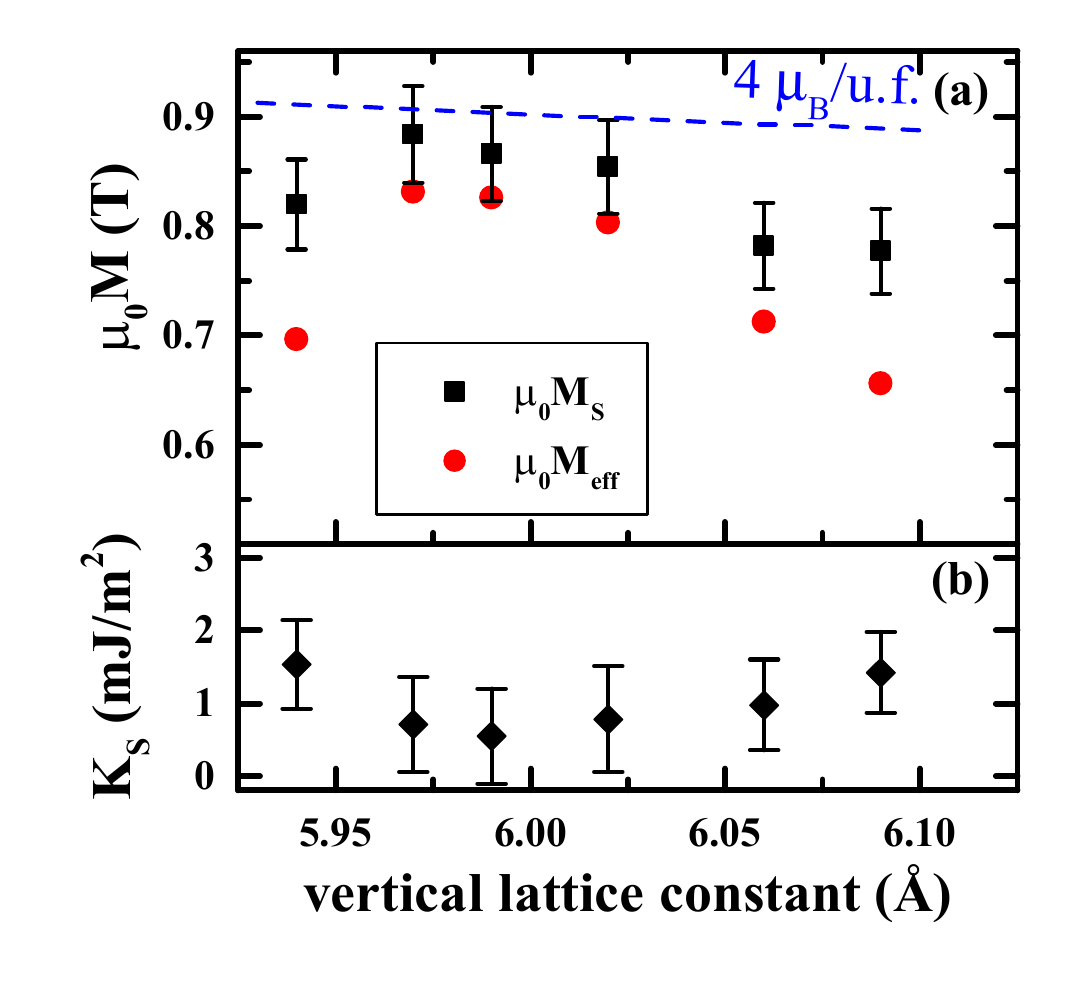}
\caption{(a) $M_S$ and $M_{\text{eff}}$ as functions of vertical lattice constant. The theoretical value of 4.0~$\mu_{B}/\text{u.f.}$ is shown by the blue dashed line. (b) The calculated surface anisotropy density follows from the difference between $M_S$ and $M_{\text{eff}}$.}\label{fig:fig2}
\end{figure}

The values of $\mu_{0} M_{\text{eff}}$ are plotted in Fig.~\ref{fig:fig2}(a) as red dots. The effective magnetization is considerably lower than the saturation magnetization $\mu_{0} M_{S}$, which was independently assessed using SQUID measurements and alternating gradient magnetometry (AGM). The values for $\mu_0 M_S$ correspond to a saturation magnetization between 3.5~$\mu_{B}/\text{unit formula}$ and 3.9~$\mu_{B}/\text{u.f.}$, with the latter value being within the error bars of the theoretically expected value of 4.0~$\mu_{B}/\text{u.f.}$ for stoichiometric NiMnSb.~\cite{graf2011pssc} A reduction of $M_{S}$ is expected in Mn-rich NiMnSb alloys, due to the antiferromagnetic coupling of the Mn$_{\text{Ni}}$ defects to the Mn lattice in the C1$_{\text{b}}$ structure of the half-Heusler material.~\cite{ekholm2010jap} An even stronger reduction is observed for the Ni-rich sample 1, which is in accordance with the formation of Ni$_{\text{Mn}}$ antisites.~\cite{alling2006prb}

While the measurement error for $M_S$ is comparatively large due to uncertainties in the volume determination, the error bars for $M_{\text{eff}}$, as obtained from ferromagnetic resonance, are negligible. NiMnSb films have been shown to possess a small but substantial perpendicular magnetic anisotropy, which can arise from either interfacial anisotropy or lattice strain.~\cite{koveshnikov2005jap,riegler2011thesis} To quantify the difference observed between $M_S$ and $M_{\text{eff}}$, we assume a uniaxial perpendicular anisotropy due to a surface anisotropy energy density $K_S$, which is known to follow the relation:~\cite{chen2007jap}
\begin{equation}
\label{eq:K_S}
    \mu_{0} M_{\text{eff}} = \mu_{0} M_{S} - \frac{2 K_{S}}{M_{S} d} {.}
\end{equation}

The $K_S$ calculated in this way has values between 0.5~$mJ/m^2$ and 1.5~$mJ/m^2$, as shown in Fig.~\ref{fig:fig2}(b); these are comparable to the surface anisotropies obtained in other crystalline thin film systems.~\cite{liu2011jap}. Although the film thicknesses in our set vary unsystematically, we can observe  systematic behavior of $K_S$ with the vertical lattice constant, with an apparent minimum under the conditions where stoichiometric NiMnSb is expected---that is, for samples 2 and 3. The increasing values for off-stoichiometric NiMnSb can be thus attributed to the concomitant increase in lattice defects, and thus of surface defects, in these films.

\subsection{Exchange Stiffness and Gilbert Damping}

\begin{figure}[t]
\includegraphics*[width=85mm]{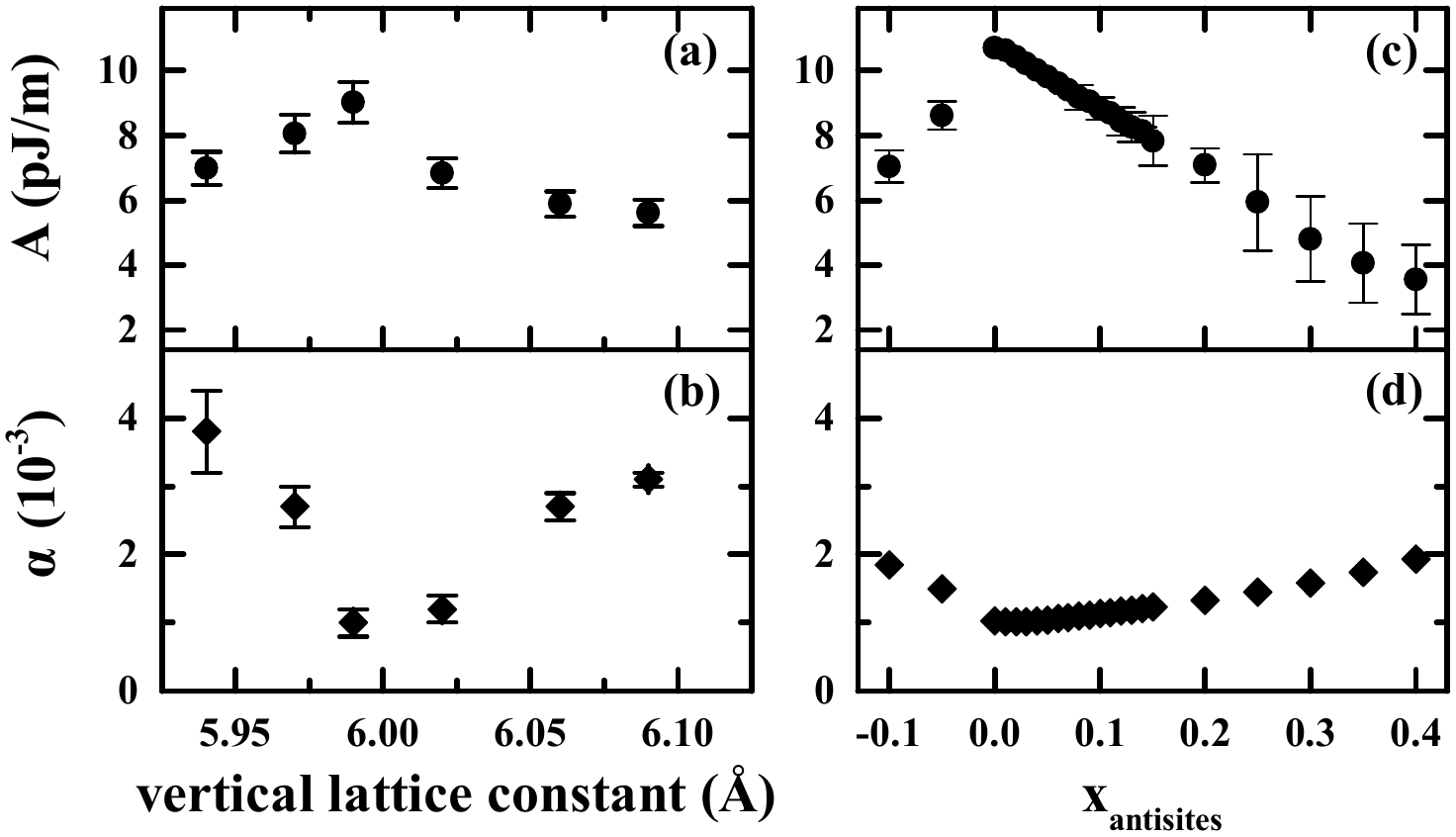}
\caption{(a) and (b) show respectively the exchange stiffness and Gilbert damping constant obtained from FMR measurements, plotted as a function of the vertical lattice constant. (c) and (d) show the corresponding values obtained from first-principle calculations for $T$~=~300~K. Negative values for $x$ imply the introduction of Ni$_{\text{Mn}}$ antisites and positive values are related to Mn$_{\text{Ni}}$ antisite defects. The error bars in (c) are the standard deviations from repeated first-principles calculations with 200 randomized supercells.}\label{fig:fig3}
\end{figure}

The experimentally determined exchange stiffness, as a function of the vertical lattice constant, and the Gilbert damping parameter are shown in Fig.~\ref{fig:fig3}(a) and (b), respectively. The minimum damping observed in our measurements is $1.0 \times 10^{-3}$ for sample 3,  and so within stoichiometric composition. Sample 1, with a deficiency of Mn atoms, showed nonlinear linewidth behavior at low frequencies, which vanished for out-of-plane measurements (not shown). This is typical with the presence of two-magnon scattering processes.~\cite{liu2011jap} However,  the damping is considerably lower in all samples than in a permalloy film of comparable thickness.

The exchange stiffness and Gilbert damping  obtained from the first-principles calculations are shown in Fig.~\ref{fig:fig3}(c) and (d), respectively. For both parameters, the experimental trends are reproduced quantitatively, with $A$ having a maximum and $\alpha$ a minimum value at stoichiometry.

As the concentration of both Mn or Ni antisites increases, the exchange stiffness decreases. This behavior can be explained by analyzing the terms in the expression for the spin-wave stiffness, Eq.~\ref{eq:exchange_stiff}. It turns out that the new exchange couplings $J_{ij}$, which appear when antisites are present, play a major role, whereas changes in the atomic magnetic moments or the saturation magnetization appear to be relatively unimportant. Mn antisites in the Ni sublattice (i.e., excess Mn) have a strong (2~mRy) antiferromagnetic coupling to the Mn atoms in the adjacent Mn layers. This results in a negative contribution to $D$ compared to the stoichiometric case, where this interaction is not present.
On the other hand, Ni antisites in the Mn sublattice have a negative in-plane exchange coupling of 0.3~mRy to their nearest-neighbor Mn atoms, with a frustrated antiferromagnetic coupling to the Ni atoms in the adjacent Ni plane. The net effect is a decreasing spin-wave stiffness as the composition moves away from stoichiometry. The calculated values of $A$ are around 30~\% larger than the experimental results, which is the same degree of overestimation we recently observed in a study of  doped permalloy films~\cite{yin2015prb}. It thus seems to be inherent in our calculations from first principles.

The calculated Gilbert damping also agrees well with the experimental values. The damping has its minimum value  of 1.0$\times$10$^{-3}$ at stoichiometry and increases with a surplus of Ni faster than with the same surplus of Mn. Both Mn and Ni antisites will act as impurities and it is thus reasonable to attribute the observed increase in damping at off-stoichiometry to impurity scattering. While the damping at stoichiometry also agrees  quantitatively, the increase in damping is underestimated in the calculations compared to the experimental values.

Despite the fact that the calculations here focus purely on the formation of Mn$_{\text{Ni}}$ or Ni$_{\text{Mn}}$ antisites, they are nonetheless capable of reproducing the experimental trends well. However,  interstitials---that is, Mn or Ni surplus atoms in the vacant sublattice---may also be a possible off-stoichiometric defect in our system.~\cite{alling2006prb} We have calculated their effects and can therefore discuss about the existence of interstitials in our samples. A large fraction of Mn interstitials seems unlikely, as an increase in the saturation magnetization can be predicted through calculations, contrary to the experimental trend; see Fig.~\ref{fig:fig2}(a). On the other hand, the existence of Ni interstitials may be compatible with the observed experimental trend, as they decrease the saturation magnetization---albeit at a slower rate than Ni antisites and slower than experimentally observed. Judging from the measured data, it is therefore likely that excess Ni exists in the samples as both antisites and interstitials.

\section{Conclusions}
In summary, we have found that off-stoichiometry in the epitaxially grown half-Heusler alloy NiMnSb has a significant impact on the material's magnetodynamic properties. In particular, the exchange stiffness can be altered by a factor of about 2 while keeping the Gilbert damping very low ($\approx$5 times lower than in permalloy films). This is a unique combination of properties and opens up for the use of NiMnSb in, \textit{e.g.}, magnonic circuits, where a small spin wave damping is desired. At the stoichiometric composition, the saturation magnetization and exchange stiffness take on their maximum values, whereas the Gilbert damping parameter is at its minimum. These experimentally observed results are reproduced by calculations from first principles. Using these calculations, we can also explain the microscopic mechanisms behind the observed trends. We also conclude that interstitial Mn is unlikely to be present in the samples. The observed effects can be used to fine-tune the magnetic properties of NiMnSb films towards their specific requirements in spintronic devices.

\begin{acknowledgments}
We acknowledge financial support from the G\"oran Gustafsson Foundation, the Swedish Research Council (VR), Energimyndigheten (STEM), the Knut and Alice Wallenberg Foundation (KAW), the Carl Tryggers Foundation (CTS), and the Swedish Foundation for Strategic Research (SSF). F.G. acknowledges financial support from the University of W\"{u}rzburg's ``Equal opportunities for women in research and teaching'' program. This work was also supported initially by the European Commission FP7 Contract ICT-257159 ``MACALO''. A.B acknowledges eSSENCE. The computer simulations were performed on resources provided by the Swedish National Infrastructure for Computing (SNIC) at the National Supercomputer Centre (NSC) and High Performance Computing Center North (HPC2N).

\end{acknowledgments}

%


\end{document}